# From Lorenz to Coulomb and other explicit gauge transformations

J. D. Jackson
Department of Physics and Lawrence Berkeley National Laboratory
University of California, Berkeley
Berkeley, CA 94720

**ABSTRACT**
The main purposes of this paper are (i) to illustrate explicitly by a number of examples the gauge functions $\chi(\mathbf{x},t)$ whose spatial and temporal derivatives transform one set of electromagnetic potentials into another equivalent set; and (ii) to show that, whatever propagation or non-propagation characteristics are exhibited by the potentials in a particular gauge, the electric and magnetic *fields* are always the same and display the experimentally verified properties of causality and propagation at the speed of light. The example of the transformation from the Lorenz gauge (retarded solutions for both scalar and vector potential) to the Coulomb gauge (instantaneous, action-at-a-distance, scalar potential) is treated in detail. A transparent expression is obtained for the vector potential in the Coulomb gauge, with a finite nonlocality in time replacing the expected spatial nonlocality of the transverse current. A class of gauges (v-gauge) is described in which the scalar potential propagates at an arbitrary speed *v* relative to the speed of light. The Lorenz and Coulomb gauges are special cases of the v-gauge. The last examples of gauges and explicit gauge transformation functions are the Hamiltonian or temporal gauge, the nonrelativistic Poincaré or multipolar gauge, and the relativistic Fock-Schwinger gauge.

## I. INTRODUCTION

The use of potentials in electromagnetism has a long history. The long path to an understanding that the vector and scalar potentials are not unique and that different potentials describing the same physics are connected by something called a gauge transformation has been described by Okun and me elsewhere.[1] If a given situation in electromagnetism is attributed to a scalar potential $\Phi(\mathbf{x},t)$ and a vector potential $\mathbf{A}(\mathbf{x},t)$, the physically meaningful and unique electric and magnetic fields $\mathbf{E}(\mathbf{x},t)$ and $\mathbf{B}(\mathbf{x},t)$ are determined from the potentials according to

$$\mathbf{E}(\mathbf{x},t) = -\nabla\Phi(\mathbf{x},t) - \frac{1}{c}\frac{\partial \mathbf{A}(\mathbf{x},t)}{\partial t}$$

$$\mathbf{B}(\mathbf{x},t) = \nabla \times \mathbf{A}(\mathbf{x},t)$$

(1.1)

Here we are using Gaussian units and considering phenomena in vacuum or as microscopic fields with localized sources. The expressions (1.1) are constituted so that they satisfy the homogeneous Maxwell equations automatically. Since the gradient of a scalar function has zero curl, it is clear that the magnetic field is unchanged if we add to $\mathbf{A}$ the gradient of a scalar function. Of course, such an addition changes the expression for the electric field. We must therefore modify the scalar potential, too. These changes are called a *gauge transformation*. Specifically, we have new potentials, $\Phi'(\mathbf{x},t)$ and $\mathbf{A}'(\mathbf{x},t)$,

$$\mathbf{A}'(\mathbf{x},t) = \mathbf{A}(\mathbf{x},t) + \nabla \chi(\mathbf{x},t)$$
$$\Phi'(\mathbf{x},t) = \Phi(\mathbf{x},t) - \frac{1}{c}\frac{\partial \chi(\mathbf{x},t)}{\partial t} \tag{1.2}$$

where the scalar function $\chi(\mathbf{x},t)$ is called the *gauge function*. The potentials $\mathbf{A}'(\mathbf{x},t)$, $\Phi'(\mathbf{x},t)$ are fully equivalent to the original set $\mathbf{A}(\mathbf{x},t)$, $\Phi(\mathbf{x},t)$, yielding the *same* electric and magnetic fields, but satisfying different dynamical equations. The chief purposes of this paper are to demonstrate some gauge transformations explicitly and to show explicitly that potentials in those different gauges, though different in detail, always yield the same electric and magnetic fields.

As is described in the textbooks[2, 3, 4], common choices of gauge are $\nabla \cdot \mathbf{A} = 0$, called the *Coulomb gauge*, and the relativistically covariant $\partial_\mu A^\mu = 0$ ( $\nabla \cdot \mathbf{A} + \frac{1}{c}\frac{\partial \Phi}{\partial t} = 0$ ), called the *Lorenz gauge*.[5] There are many other gauges, but the textbooks rarely show explicitly the gauge function $\chi$ that transforms one gauge into another.

In Section II we review the form of the equations and the solutions for the potentials in the Lorenz gauge. We also exhibit the corresponding equations in the Coulomb gauge, focusing on the nonlocality of the source for the vector potential. The direct solution is deferred to Section IV. In Section III the gauge function $\chi(\mathbf{x},t)$ to go from the Lorenz gauge to the Coulomb gauge is constructed and used to calculate the Coulomb-gauge vector potential. The results (3.10) and (3.16) or (3.17), are surprisingly explicit and compact, with only one time integral over a finite range of the source's time $t'$ ($t - R/c < t' < t$), replacing the spatial nonlocality of the source with a temporal nonlocality. We return to the original equation for the Coulomb-gauge vector potential in Section IV and show that its straightforward solution can be transformed into that obtained in Section III more directly and simply through the gauge function.

In Section V we derive the electric and magnetic *fields* from the Coulomb-gauge potentials and show that they are the well known expressions, causal and propagating with speed *c*, despite the instantaneous nature of the scalar potential. This ground has been traveled before in this journal by Brill and Goodman[6] and recently by Rohrlich.[7] There is also Problem 6.20 in my book.[2] Our discussion here is different and I think more transparent because of the form of our solution for $\mathbf{A}_C$. Some aspects of Brill and Goodman come close. In Section VI we discuss briefly the quasi-static limit of the vector potential in the Coulomb gauge and its use to obtain a Lagrangian for the interaction of charged particles that is correct to order $1/c^2$ in the velocities. Section VII is devoted to a class of gauges we call the velocity gauge (v-gauge) in which the scalar potential propagates with an *arbitrary speed v*. The Lorenz and Coulomb gauges are limiting cases, $v = c$ and $v = \infty$, respectively. The gauge function and the potentials are determined, as are the electromagnetic fields (the same as always).

The Hamiltonian or temporal gauge, with its vanishing scalar potential, is treated in Section VIII, while Section IX describes the nonrelativistic and relativistic gauges that go under the names of Poincaré, multipolar, Fock, or Schwinger. The gauge functions are exhibited as well as the potentials. For the nonrelativistic case the interaction Lagrangian in the multipolar gauge is described. Finally, a summary is given in Section X. An appendix presents the Lorenz-to-Coulomb gauge function in terms of the current density rather than the charge density and finds its quasi-static limit.

## II. POTENTIALS IN THE LORENZ GAUGE AND SOURCES FOR THE COULOMB GAUGE

Traditionally, the solutions for the potentials are presented first in the Lorenz gauge.[8] One generally begins with the relations (1.1) substituted into the inhomogeneous Maxwell equations to give two coupled partial differential equations,

$$\nabla^2 \Phi + \frac{1}{c} \frac{\partial}{\partial t} ( \nabla \cdot \mathbf{A} ) = - 4\pi \rho \tag{2.1}$$

$$\nabla^2 \mathbf{A} - \frac{1}{c^2} \frac{\partial^2 \mathbf{A}}{\partial t^2} - \nabla \left( \nabla \cdot \mathbf{A} + \frac{1}{c} \frac{\partial \Phi}{\partial t} \right) = - \frac{4\pi}{c} \mathbf{J} \tag{2.2}$$

The freedom implied by (1.2) means that we may choose[9] (among other choices) to set

$$\nabla \cdot \mathbf{A} + \frac{1}{c} \frac{\partial \Phi}{\partial t} = 0 \quad \text{or} \quad \nabla \cdot \mathbf{A} = 0. \tag{2.3 a,b}$$

The first choice, called the Lorenz gauge, leads to the decoupled wave equations,

$$\nabla^2 \Phi_L - \frac{1}{c^2} \frac{\partial^2 \Phi_L}{\partial t^2} = - 4\pi \rho$$

$$\nabla^2 \mathbf{A}_L - \frac{1}{c^2} \frac{\partial^2 \mathbf{A}_L}{\partial t^2} = - \frac{4\pi}{c} \mathbf{J} \tag{2.4}$$

with so-called retarded solutions,

$$\Phi_L(\mathbf{x},t) = \int d^3x' \frac{1}{R} \left[ \rho(\mathbf{x}',t'=t - R/c) \right]$$

$$\mathbf{A}_L(\mathbf{x},t) = \frac{1}{c} \int d^3x' \frac{1}{R} \left[ \mathbf{J}(\mathbf{x}',t'=t - R/c) \right] \tag{2.5}$$

Here $R = |\mathbf{x} - \mathbf{x}'|$. The time $t' = t - R/c$ is called the retarded time.

An alternative is to choose the second relation in (2.3), $\nabla \cdot \mathbf{A} = 0$. This choice, called the Coulomb or radiation or transverse gauge, makes the equation for the scalar potential simple, namely,

$$\nabla^2 \Phi_C = -4\pi \rho$$

with the solution,

$$\Phi_C(\mathbf{x},t) = \int d^3x' \frac{1}{R} \rho(\mathbf{x}',t) \tag{2.6}$$

This solution is called the instantaneous scalar potential because it is a solution of the Poisson equation without retardation. On the other hand, the equation for the vector potential is more complicated:

$$\nabla^2 \mathbf{A}_C - \frac{1}{c^2} \frac{\partial^2 \mathbf{A}_C}{\partial t^2} = -\frac{4\pi}{c} \mathbf{J} - \frac{1}{4\pi} \nabla \frac{\partial \Phi_C}{\partial t} \tag{2.7}$$

The effective current on the right hand side is known as the transverse current. It has zero divergence. Recall that any vector field can be decomposed into a sum of a longitudinal or irrotational part (with zero curl) and a transverse or solenoidal part (with zero divergence). We can write the total current to show explicitly the two parts:[9]

$$\mathbf{J} = \mathbf{J}_l + \mathbf{J}_t$$

with

$$\mathbf{J}_l = -\frac{1}{4\pi} \nabla \int d^3x' \frac{1}{R} \nabla' \cdot \mathbf{J}(\mathbf{x}',t) = \frac{1}{4\pi} \nabla \int d^3x' \frac{1}{R} \frac{\partial}{\partial t} \rho(\mathbf{x}',t)$$

$$\mathbf{J}_t = \frac{1}{4\pi} \nabla \times \nabla \times \int d^3x' \frac{1}{R} \mathbf{J}(\mathbf{x}',t) \tag{2.8}$$

Note that the second form of $\mathbf{J}_l$ is just the negative of the second term in (2.7) The source term on the right in (2.7) is therefore $\mathbf{J}_t$. A vector potential can be decomposed in the same way as the current. We write

$$\mathbf{A}_l = \nabla \Psi \; ; \qquad \mathbf{A}_t = \nabla \times \mathbf{V} \tag{2.9}$$

For the Coulomb-gauge vector potential, the auxiliary functions $\Psi$ and $\mathbf{V}$ satisfy inhomogeneous wave equations,

$$\nabla^2 \Psi - \frac{1}{c^2}\frac{\partial^2 \Psi}{\partial t^2} = -\frac{1}{c}\int d^3x' \frac{1}{R}\frac{\partial}{\partial t}\rho(\mathbf{x}',t)$$

$$\nabla^2 \mathbf{V} - \frac{1}{c^2}\frac{\partial^2 \mathbf{V}}{\partial t^2} = -\frac{1}{c}\nabla \times \int d^3x' \frac{1}{R}\mathbf{J}(\mathbf{x}',t)$$

(2.10)

Since the source terms are nonlocal in space, the retarded solutions ostensibly involve two 3-dimensional spatial integrals. In the next section, we find the gauge function $\chi$ that takes us from the Lorenz gauge to the Coulomb gauge. We show that in fact the solution in the Coulomb gauge needs only one 3-dimensional spatial integration plus a time integration over a finite interval (nonlocality in time). That the two forms of the solution are equivalent is demonstrated in Section IV.

### III. GAUGE FUNCTION FOR TRANSFORMATION FROM THE LORENZ TO THE COULOMB GAUGE, VECTOR POTENTIAL IN COULOMB GAUGE

**A. Gauge function in terms of the charge density**

To find the gauge function $\chi_C$ we focus on the two scalar potentials. The scalar potential in the Lorenz gauge is

$$\Phi_L(\mathbf{x},t) = \int d^3x' \frac{1}{R}\rho(\mathbf{x}',t'=t-R/c) \tag{3.1}$$

In the Coulomb gauge the instantaneous scalar potential reads

$$\Phi_C(\mathbf{x},t) = \int d^3x' \frac{1}{R}\rho(\mathbf{x}',t) \tag{3.2}$$

Equation (1.2) tells us that the partial time derivative of the gauge function $\chi_C$ that transforms the Lorenz scalar potential into the Coulomb scalar potential is

$$\frac{1}{c}\frac{\partial \chi_C}{\partial t} = \Phi_L(\mathbf{x},t) - \Phi_C(\mathbf{x},t) = \int d^3x' \frac{1}{R}\left[\rho(\mathbf{x}',t'=t-R/c) - \rho(\mathbf{x}',t)\right] \tag{3.3}$$

We integrate both sides with respect to $ct$ to obtain

$$\chi_C(\mathbf{x},t) = c\int d^3x' \frac{1}{R}\left[\int_{t_0}^{t-R/c} dt' \rho(\mathbf{x}',t') - \int_{t_0}^{t} dt' \rho(\mathbf{x}',t')\right] + \chi_0 \tag{3.4}$$

This can be written more compactly as

$$\chi_C(\mathbf{x},t) = c \int d^3x' \frac{1}{R} \int_t^{t-R/c} dt'\, \rho(\mathbf{x}',t') + \chi_0 \tag{3.5}$$

If we change variables by writing $t' = t - \tau$, then we have

$$\chi_C(\mathbf{x},t) = -c \int d^3x' \frac{1}{R} \int_0^{R/c} d\tau\, \rho(\mathbf{x}',t-\tau) + \chi_0 \tag{3.6}$$

A somewhat more compact form is

$$\chi_C = -\int d^3x' \int_0^1 d\lambda\, \rho(\mathbf{x}',t - \lambda R/c) + \chi_0 \tag{3.7}$$

[In constructing the vector potential the first form (3.6) is actually easier to use.] Equation (3.6) or (3.7) is an explicit expression for the gauge function $\chi_C$ in terms of the charge density. The integration term $\chi_0$ is a priori a function of $\mathbf{x}$, but not of time. In fact, it is at most a constant, as can be seen as follows. The requirement that $\nabla \cdot \mathbf{A}_C = 0$ leads to a wave equation for $\chi_C$,

$$\nabla^2 \chi_C - \frac{1}{c^2} \frac{\partial^2 \chi_C}{\partial t^2} = \frac{1}{c} \frac{\partial \Phi_C}{\partial t} \tag{3.8}$$

The term involving the charge density in (3.6) is the particular integral of this equation. The term $\chi_0$ represents the addition of a solution to the homogeneous equation. But such a solution can consist only of plane waves, with time dependence. A time-independent $\chi_0$, evidently a solution of the Laplace equation, can at most be a constant if we demand finiteness at infinity. In passing, we note the closeness of (3.8) to the equation (2.10) for the auxiliary function $\Psi$. In Section IV we show how to get from (3.8) or (2.10) to the form (3.6).

Characteristically (3.6) contains values of the charge density at all times from the retarded time ($t-R/c$) to the present time $t$. An equivalent expression for $\chi_C$ involving the current density $\mathbf{J}(\mathbf{x}',t-\tau)$ is derived in Appendix A.

## B. Vector potential in Coulomb gauge

Since we know the scalar potential in the Coulomb gauge (3.2) all that remains is to add the gradient of (3.6) to the vector potential from (2.5) to find the vector potential in this gauge. The gradient of (3.6) is

$$\chi_C = -c \int d^3x' \frac{1}{R} \int_0^{R/c} d\tau\, \rho(\mathbf{x}', t-\tau)$$

The brackets show the terms that are to be differentiated by the gradient operator. The result of the operation is

$$\chi_C = \int d^3x' \left[ -\frac{\hat{\mathbf{R}}}{R} \rho(\mathbf{x}', t - R/c) + \frac{c\hat{\mathbf{R}}}{R^2} \int_0^{R/c} d\tau\, \rho(\mathbf{x}', t-\tau) \right] \quad (3.9)$$

Addition of (3.9) to (2.5) yields the Coulomb-gauge vector potential:

$$\mathbf{A}_C(\mathbf{x}, t) = \frac{1}{c} \int d^3x' \frac{1}{R} \left[ \mathbf{J}(\mathbf{x}', t') - c\hat{\mathbf{R}} \rho(\mathbf{x}', t') \right]_{ret} + \frac{c^2 \hat{\mathbf{R}}}{R^2} \int_0^{R/c} d\tau\, \rho(\mathbf{x}', t-\tau) \quad (3.10)$$

Here we have utilized the abbreviation "*ret*" to indicate that the quantity inside the square brackets is to be evaluated at the retarded time $t' = t - R/c$. The first part of the vector potential is a standard retarded form. The additional time integral over all times $t'$ from $t' = t - R/c$ to $t' = t$ is the price we pay for having the simple instantaneous scalar potential. The price is, however, lower than appeared initially from the spatial nonlocality of the sources. We show in Section V how the *fields* calculated from the potentials in the Coulomb gauge are the same as those found from the Lorenz gauge (and any other gauge). fully obeying causality and a finite speed of propagation.

    An alternative expression for the Coulomb-gauge vector potential, totally in terms of the current density, can be obtained as follows. The two terms in (3.9) can be combined by an integration by parts:

$$\chi_C = -c \int d^3x' \frac{\hat{\mathbf{R}}}{R^2} \int_0^{R/c} d\tau\, \tau \frac{\partial}{\partial \tau} \rho(\mathbf{x}', t-\tau) \quad (3.11)$$

With $\frac{\partial \rho}{\partial t} = -\frac{\partial \rho}{\partial \tau}$, we can use the continuity equation for charge and current densities to write (3.11) in terms of the divergence of the current density:

$$\chi_C = -c \int d^3x' \frac{\hat{\mathbf{R}}}{R^2} \int_0^{R/c} d\tau\, \tau\, \nabla' \cdot \mathbf{J}(\mathbf{x}', t-\tau) \quad (3.12)$$

An integration by parts throws the gradient operator $\nabla'$ over onto the factors of $\mathbf{R}$. Then with $\nabla' f(\mathbf{R}, R) = -\nabla f(\mathbf{R}, R)$, we find

$$\chi_C = -c \int d^3x' \frac{\partial}{\partial x_k} \frac{\hat{\mathbf{R}}}{R^2} \int_0^{R/c} d\tau\, \tau\, J_k(\mathbf{x}', t-\tau) \qquad (3.13)$$

The differentiation acts only on the $\mathbf{R}$-dependence. The components of $\mathbf{J}$ are fixed. For clarity consider the $j^{th}$ component of the gradient. We need

$$\frac{\partial}{\partial x_k} \frac{R_j}{R^3} J_k = \frac{1}{R^3}\left(J_j - 3\hat{R}_j(\hat{\mathbf{R}}\cdot\mathbf{J})\right) + \frac{4\pi}{3} J_j \delta^{(3)}(\mathbf{R}) \qquad (3.14)$$

The delta function term is analogous to that appearing in the electric field of a point dipole.[10] Because it appears under the integral sign and corresponds to $R = 0$ it does not contribute. The other term we need is

$$\frac{\hat{R}_j}{R^2} \frac{\partial}{\partial x_k} \int_0^{R/c} d\tau\, \tau\, J_k = \frac{1}{c^2 R} \hat{R}_j \left[\hat{\mathbf{R}}\cdot\mathbf{J}(\mathbf{x}', t')\right]_{ret} \qquad (3.15)$$

If we insert (3.14) and (3.15) into (3.13) and add to (2.5) we find an alternative form for the Coulomb-gauge vector potential entirely in terms of $\mathbf{J}$:

$$\begin{aligned}
\mathbf{A}_C &= \frac{1}{c} \int d^3x' \frac{1}{R}\left[\mathbf{J}(\mathbf{x}', t') - \hat{\mathbf{R}}\,\hat{\mathbf{R}}\cdot\mathbf{J}(\mathbf{x}', t')\right]_{ret} \\
&\quad + c \int d^3x' \frac{1}{R^3} \int_0^{R/c} d\tau\, \tau \left[3\hat{\mathbf{R}}\,\hat{\mathbf{R}}\cdot\mathbf{J}(\mathbf{x}', t-\tau) - \mathbf{J}(\mathbf{x}', t-\tau)\right]
\end{aligned} \qquad (3.16)$$

The first line is the contribution from the retarded and locally transverse current. It is the only term that survives in the radiation zone of a localized source. The second line, with its more rapid decrease with distance contributes to the potential in the near and intermediate zones.

It turns out that the rather cumbersome (3.16) can be written compactly as a curl (as it must - see (2.9)):

$$\mathbf{A}_C(\mathbf{x}, t) = -c\,\nabla\times \int d^3x' \int_0^{R/c} d\tau\, \tau\, \frac{\hat{\mathbf{R}}\times\mathbf{J}(\mathbf{x}', t-\tau)}{R^2} \qquad (3.17)$$

Either this compact form or the more explicit (3.16) together with the instantaneous scalar potential (3.2), repeated here,

$$\Phi_C(\mathbf{x},t) = \int d^3x' \frac{1}{R} \rho(\mathbf{x}',t) \tag{3.2}$$

are the electromagnetic potentials in the Coulomb gauge. Equation (3.16) is particularly useful in discussing the quasi-static limit - see Section VI. In passing we note that the path from (3.11) to (3.17) can be replaced by analogous steps that begin with the current density form of $\chi_C$, equation (A.4) of the Appendix.

## IV. DIRECT SOLUTION OF THE WAVE EQUATION FOR THE VECTOR POTENTIAL IN THE COULOMB GAUGE

The previous section obtained the Coulomb-gauge vector potential by constructing the gauge function $\chi_C$. Now we solve the wave equations (2.10) for the auxiliary functions $\Psi$ and **V**. The retarded solution for $\Psi$ is evidently

$$\Psi(\mathbf{x},t) = \frac{1}{4\pi c} \int d^3x'' \frac{1}{R'} \int d^3x' \frac{1}{R''} \frac{\partial}{\partial t'} \rho(\mathbf{x}',t') \bigg|_{ret} \tag{4.1}$$

Here $\mathbf{R}' = \mathbf{x} - \mathbf{x}''$, while $\mathbf{R}'' = \mathbf{x}'' - \mathbf{x}'$ and "ret" means $t' = t - R'/c$ (NOT $t' = t - R''/c$). The somewhat peculiar choice of primes is with malice aforethought. Our plan of action is to find $\Psi$ as an integral over three spatial dimensions and one time dimension, as in (3.17) for example, then take its gradient as in (2.9) to find the longitudinal vector potential, and subtract that from the Lorenz-gauge vector potential (2.5) to find the transverse (Coulomb-gauge) vector potential.

We proceed by interchanging the orders of integration in (4.1), We need the familiar coordinate $\mathbf{R} = \mathbf{x} - \mathbf{x}' = \mathbf{R}' + \mathbf{R}''$. Then we have $R'' = |\mathbf{R} - \mathbf{R}'|$. As we interchange the orders of integration we also shift the origin so that $d^3x'' \to d^3R'$. Then we have

$$\Psi(\mathbf{x},t) = \frac{1}{4\pi c} \int d^3x' \int d^3R' \frac{1}{R'|\mathbf{R}' - \mathbf{R}|} \frac{\partial}{\partial t'} \rho(\mathbf{x}',t - R'/c) \tag{4.2}$$

We observe that in the angular part of the $d^3R'$ integration the charge density factor is constant, only depending on the magnitude of $R'$. The inverse distance $1/|\mathbf{R}' - \mathbf{R}|$ can be expanded in spherical harmonics,

$$\frac{1}{|\mathbf{R}' - \mathbf{R}|} = \sum_{l=0}^{\infty} \frac{r_<^l}{r_>^{l+1}} P_l(\hat{\mathbf{R}} \cdot \hat{\mathbf{R}}')$$

In the integration over angles of $\mathbf{R}'$ only the $l = 0$ term survives. Thus (4.2) becomes

$$\Psi(\mathbf{x},t) = - \int d^3x' \int_0^\infty dR' R' \frac{1}{r_>} \frac{\partial}{\partial R'} \rho(\mathbf{x}',t-R'/c) \tag{4.3}$$

Here we have used the fact that for $t$ fixed $\frac{\partial \rho}{\partial t'} = -c \frac{\partial \rho}{\partial R'}$. The variable $r_>$ is the larger of $R'$ and $R$. The integral over $dR$ is broken into two parts:

$$\int_0^\infty dR' R' \frac{1}{r_>} \frac{\partial}{\partial R'} \rho = \frac{1}{R} \int_0^R dR' R' \frac{\partial \rho}{\partial R'} + \int_R^\infty dR' \frac{\partial \rho}{\partial R'}$$

Integration by parts on the first term leads to

$$\int_0^\infty dR' R' \frac{1}{r_>} \frac{\partial}{\partial R'} \rho = \frac{1}{R} \left( R'\rho \Big|_0^R - \int_0^R dR' \rho(\mathbf{x},t-R'/c) \right)$$

$$+ \lim_{\tau} \rho(\mathbf{x}',t-\tau) - \rho(\mathbf{x}',t-R/c)$$

The first term in the large curved brackets is canceled by the last term in the second line. The first term in the second line is the charge density at a remote time in the past. Its volume integral in (4.3) is equal to the total charge $Q$, assumed constant. If we replace $dR'$ with $cd\tau$, (4.3) becomes

$$\psi(\mathbf{x},t) = \int d^3x' \frac{c}{R} \int_0^{R/c} d\tau\, \rho(\mathbf{x}',t-\tau) - Q \tag{4.4}$$

As was hinted at below (3.8), comparison of (4.4) with the gauge function (3.6) shows that

$$\Psi(\mathbf{x},t) = \chi_0 - Q - \chi_C(\mathbf{x},t) \tag{4.5}$$

Since the longitudinal vector potential is $\mathbf{A}_l = \nabla \Psi = -\nabla \chi_C$ and the Coulomb-gauge vector potential is the transverse vector potential, we find

$$\mathbf{A}_C = \mathbf{A}_L - \mathbf{A}_l = \mathbf{A}_L + \nabla \chi_C \tag{4.6}$$

We have thus shown that the straightforward solution of the wave equations (2.10) leads to the same result as found in Section III by means of $\chi_C$. The explicit construction of the Coulomb-gauge vector potential in the form (3.16) or (3.17) can also be found by solving the second equation in (2.10) for **V**, although more work is required.

## V. ELECTRIC AND MAGNETIC FIELDS FROM THE COULOMB-GAUGE POTENTIALS

### A. Fields obtained from the Coulomb-gauge potentials

It seems necessary from time to time to show that the electric and magnetic *fields* are independent of the choice of gauge for the potentials. In particular, the Coulomb gauge, with its instantaneous scalar potential, seems to contradict the ingrained truth that all electromagnetic fields in vacuum propagate with the speed of light. In this section we focus on that troublesome gauge, taking as our starting point the potentials (3.2) and (3.9),

$$\mathbf{A}_C(\mathbf{x},t) = \frac{1}{c}\int d^3x' \frac{1}{R}\left[\mathbf{J}(\mathbf{x}',t') - c\hat{\mathbf{R}}\rho(\mathbf{x}',t')\right]_{ret} + \frac{c^2\hat{\mathbf{R}}}{R^2}\int_0^{R/c} d\tau\, \rho(\mathbf{x}',t-\tau)$$

$$\Phi_C(\mathbf{x},t) = \int d^3x' \frac{1}{R}\rho(\mathbf{x}',t)$$
(5.1)

The electric field is $\mathbf{E} = -\nabla\Phi_C - \frac{1}{c}\frac{\partial \mathbf{A}_C}{\partial t}$. Consider the time derivative of the vector potential,

$$-\frac{1}{c}\frac{\partial}{\partial t}\mathbf{A}_C(\mathbf{x},t) = -\frac{1}{c^2}\int d^3x' \frac{1}{R}\left[\frac{\partial}{\partial t'}\mathbf{J}(\mathbf{x}',t') - c\hat{\mathbf{R}}\frac{\partial}{\partial t'}\rho(\mathbf{x}',t')\right]_{ret} - \frac{c^2\hat{\mathbf{R}}}{R^2}\int_0^{R/c} d\tau\, \frac{\partial}{\partial \tau}\rho(\mathbf{x}',t-\tau)$$

Here we have used the fact that we can move the time derivation through the retarded bracket[11] and that $\partial/\partial t = -\partial/\partial \tau$. The last term can be integrated trivially to give

$$-\frac{1}{c}\frac{\partial}{\partial t}\mathbf{A}_C(\mathbf{x},t) = -\frac{1}{c^2}\int d^3x' \frac{1}{R}\left[\frac{\partial}{\partial t'}\mathbf{J}(\mathbf{x}',t') - c\hat{\mathbf{R}}\frac{\partial}{\partial t'}\rho(\mathbf{x}',t') - \frac{c^2\hat{\mathbf{R}}}{R^2}\rho(\mathbf{x}',t')\right]_{ret} + \frac{c^2\hat{\mathbf{R}}}{R^2}\rho(\mathbf{x}',t)$$

The first three terms here are retarded contributions to the electric field. The last term is an instantaneous contribution. A moment's thought shows that it is the positive gradient of the scalar potential in (5.1). This piece is the part of the Coulomb-gauge vector potential that cancels the instantaneous piece of the electric field from the scalar potential. The net result for the electric field is

$$\mathbf{E}(\mathbf{x},t) = \left. \int d^3x' \left[ \frac{\hat{\mathbf{R}}}{R^2} \rho(\mathbf{x}',t') + \frac{\hat{\mathbf{R}}}{cR} \frac{\partial}{\partial t'} \rho(\mathbf{x}',t') - \frac{1}{c^2 R} \frac{\partial}{\partial t'} \mathbf{J}(\mathbf{x}',t') \right] \right|_{ret} \tag{5.2}$$

The magnetic field can similarly be found to be

$$\mathbf{B}(\mathbf{x},t) = \frac{1}{c} \int d^3x' \left. \left[ \frac{1}{R^2} \mathbf{J}(\mathbf{x}',t') + \frac{1}{cR} \frac{\partial}{\partial t'} \mathbf{J}(\mathbf{x}',t') \right] \right|_{ret} \times \hat{\mathbf{R}} \tag{5.3}$$

The terms in $\mathbf{A}_C$ (5.1) involving the charge density are proportional to $\hat{\mathbf{R}}$ times scalar functions and so have zero curl and do not contribute to $\mathbf{B}$.

Equations (5.2) and (5.3) are the "Jefimenko" expressions for the electromagnetic fields.[11] They are normally evaluated from the Lorenz-gauge potentials, but we see that the Coulomb-gauge potentials serve just as well. And any other gauge, too. They only involve retarded expressions despite the presence of other terms in the potentials. Given the completely general results in the Introduction about gauge transformations and different gauges being equivalent, it could hardly be otherwise. Nonetheless it is satisfying to see explicitly how it comes about.

**B. Harmonic time dependence**

Another instructive way to see how the Coulomb-gauge potentials conspire to produce properly retarded fields is to consider harmonic time dependence. Then the integral over time in (5.1) can be done explicitly. We assume $\rho(\mathbf{x},t) = \rho(\mathbf{x}) e^{-i\omega t}$, $\mathbf{J}(\mathbf{x},t) = \mathbf{J}(\mathbf{x}) e^{-i\omega t}$. Then the vector potential in (5.1) becomes

$$\mathbf{A}_C(\mathbf{x},t) = \frac{e^{-i\omega t}}{c} \int d^3x' \left\{ \frac{e^{ikR}}{R} \left[ \mathbf{J}(\mathbf{x}') - c\hat{\mathbf{R}} \rho(\mathbf{x}') \right] + \frac{c^2 \hat{\mathbf{R}}}{R^2} \rho(\mathbf{x}') \int_0^{R/c} d\tau\, e^{i\omega \tau} \right\} \tag{5.4}$$

Here $k = \omega/c$. After doing the time integral we have

$$\mathbf{A}_C(\mathbf{x},t) = \frac{e^{-i\omega t}}{c} \int d^3x' \left[ \frac{e^{ik R}}{R} \mathbf{J}(\mathbf{x}') + \frac{c\hat{\mathbf{R}}}{R} \rho(\mathbf{x}') \frac{(1 - ik R)e^{ik R} - 1}{i kR} \right] \quad (5.5)$$

Equation (5.5) can be separated into its instantaneous and retarded parts:

$$\mathbf{A}_C(\mathbf{x},t) = \mathbf{A}_C^{ret}(\mathbf{x},t) + \mathbf{A}_C^{inst}(\mathbf{x},t)$$

where

$$\mathbf{A}_C^{ret}(\mathbf{x},t) = \frac{e^{-i\omega t}}{c} \int d^3x' \left[ \frac{e^{ik R}}{R} \mathbf{J}(\mathbf{x}') + \frac{c\hat{\mathbf{R}}}{R} \rho(\mathbf{x}') \frac{(1 - ik R)}{ik R} \right]$$

$$\mathbf{A}_C^{inst}(\mathbf{x},t) = \frac{ic\, e^{-i\omega t}}{\omega} \int d^3x' \frac{\hat{\mathbf{R}}}{R^2} \rho(\mathbf{x}') \quad (5.6)$$

The overall factor of $e^{ikR}$ in the integrand of the first equation guarantees the retarded behavior. Now consider the contribution of the instantaneous part to the electric field,

$$-\frac{1}{c}\frac{\partial \mathbf{A}_C^{inst}}{\partial t} = -e^{-i\omega t} \int d^3x' \frac{\hat{\mathbf{R}}}{R^2} \rho(\mathbf{x}') = +e^{-i\omega t} \nabla \int d^3x' \frac{1}{R} \rho(\mathbf{x}') = \nabla \Phi_C \quad (5.7)$$

The instantaneous contribution to the electric field is therefore

$$\mathbf{E}_C^{inst} = -\nabla \Phi_C - \frac{1}{c}\frac{\partial \mathbf{A}_C^{inst}}{\partial t} = -\nabla \Phi_C + \nabla \Phi_C = 0 \quad (5.8)$$

The harmonic version of the retarded electric field (5.2) is therefore given entirely by the retarded part $\mathbf{A}_C^{ret}$ from (5.6), as is the harmonic magnetic field. We see again how the scalar and vector potentials "conspire" to give the correct causal behavior with propagation at speed *c*.

## VI. QUASI-STATIC LIMIT OF THE COULOMB-GAUGE VECTOR POTENTIAL AND THE DARWIN LAGRANGIAN
### A. The quasi-static vector potential in Coulomb gauge

The quasi-static limit is defined as the neglect of retardation. For the Lorenz gauge potentials (2.5) the structure is unchanged, one merely replaces the retarded source terms by their values at the present (observer's) time *t*. In the Coulomb gauge, the scalar potential is already

"quasi-static," i.e., instantaneous. For the vector potential (3.16) or (3.17) we merely replace the current density according to $\mathbf{J}(\mathbf{x}',t-\tau) \to \mathbf{J}(\mathbf{x}',t)$ and perform the trivial time integral to find from either expression the quasi-static vector potential in the Coulomb gauge,

$$\mathbf{A}_C^{qs} = \frac{1}{2c} \int d^3x' \frac{1}{R} \left[ \mathbf{J}(\mathbf{x}',t) + \hat{\mathbf{R}}(\hat{\mathbf{R}} \cdot \mathbf{J}(\mathbf{x}',t)) \right] \tag{6.1}$$

It is of interest to note that, as described in section (b) of the Appendix, this form of the vector potential was first written down in 1870 by Helmholtz in his discussion of the equivalence of the so-called Neumann and Weber forms for the vector potential.[12] He obtained the quasi-static function (A.5) whose gradient transformed the Lorenz gauge vector potential to this quasi-static Coulomb-gauge vector potential, the arithmetic average of the Neumann and Weber forms of the vector potential, although he of course did not use any of the language of gauges and gauge transformations. As discussed in Ref. 1, Helmholtz somewhat inaccurately says that (6.1) leads to Maxwell's theory because $\nabla \cdot \mathbf{A}_C = 0$ holds here and $\nabla \cdot \mathbf{A} = 0$ was invariably Maxwell's choice. But Maxwell never wrote down or used (6.1). We contact Helmholtz's work again in Section VII.

**B. Heaviside-Darwin Lagrangian**

In the Coulomb gauge the exact scalar potential is the instantaneous Coulomb potential. The Coulomb gauge is therefore very convenient in the treatment of interacting charged particles by Lagrangian methods where all particles have a common time. To include velocity dependent effects, the vector potentials of the moving charges must enter. To the inclusive order of $1/c^2$ effects, however, one can neglect retardation because the $\mathbf{v} \cdot \mathbf{A}/c$ terms in the Lagrangian are already of order $1/c^2$; retardation corrections contribute only to higher order in $1/c$. The vector potential (6.1) can be used.

Consider the interaction of two point charged particles with charges $q$, $q'$ and coordinates and velocities $\mathbf{x}$, $\mathbf{v}$, and $\mathbf{x}'$, $\mathbf{v}'$. Let $\mathbf{r} = \mathbf{x} - \mathbf{x}'$ and $r = |\mathbf{x} - \mathbf{x}'|$. Then the vector potential caused by charge $q'$, evaluated at the position of charge $q$ is

$$\mathbf{A}'(\mathbf{x},t) = \frac{q'}{2c\,r} \left[ \mathbf{v}' + \hat{\mathbf{r}}(\hat{\mathbf{r}} \cdot \mathbf{v}') \right] \tag{6.2}$$

The vector potential of charge $q$ evaluated at the position of $q'$ has the same form, with the interchange of primed and unprimed quantities. The scalar products of these vector potentials with the other particle's velocity times charge are evidently the same and one half their sum is equal to one term. Together with the instantaneous Coulomb interaction, the interaction Lagrangian, correct to order $1/c^2$ inclusive, is thus

$$L_{int} = -\frac{qq'}{r} + \frac{1}{2c}\left[q\mathbf{v}\cdot\mathbf{A}'(\mathbf{x},t) + q'\mathbf{v}'\cdot\mathbf{A}(\mathbf{x}',t)\right]$$
$$= \frac{qq'}{r}\left[-1 + \frac{1}{2c^2}\left(\mathbf{v}\cdot\mathbf{v}' + (\hat{\mathbf{r}}\cdot\mathbf{v})(\hat{\mathbf{r}}\cdot\mathbf{v}')\right)\right] \quad (6.3)$$

This interaction Lagrangian is called the Darwin Lagrangian, after C. G. Darwin who derived it in 1920 by expansion of the Liénard-Wiechert potentials[13], although it was derived in effect by Oliver Heaviside in 1889.[14] It has uses in plasma physics and in quantum mechanics (as the basis for the so-called Breit interaction of two electrons in atoms).

If higher order terms beyond (6.3) are desired, one may expand the current densities in (3.16) in Taylor series around $t' = t$. These terms will contain the particles' accelerations and higher derivatives with respect to time. To get a conventional Lagrangian it will then be necessary to use lower order equations of motion to eliminate terms beyond the velocities.

**VII. GAUGE FUNCTION AND POTENTIALS FOR THE VELOCITY GAUGE**

The Lorenz gauge and the Coulomb gauge are in some sense the extremes in behavior of the scalar potential. In the one the effects of a source point propagate with the speed $c$; in the other the propagation is at infinite speed. But there is no reason for a restriction to such limiting cases. We define the *velocity gauge* ( v-gauge for short and actually a class of gauges) to be one in which the speed of propagation for the scalar potential is $v$, *an arbitrary speed* relative to $c$. The velocity gauge has been discussed in classical electrodynamics by Yang,[15] Brown and Crothers,[16] and Drury,[17] and in quantum electrodynamics by Baxter,[18] with the names "$\alpha$-Lorentz gauge,"[15,16,17] and "velocity gauge."[17] Brown and Crothers are most explicit; Drury discusses the equations but does not solve them.

The gauge condition for such a gauge generalizes those of (2.3), namely,

$$\nabla\cdot\mathbf{A}_v + \alpha\frac{1}{c}\frac{\partial\Phi_v}{\partial t} = 0; \quad \alpha = \frac{c^2}{v^2} \quad (7.1)$$

The wave equation satisfied by the scalar potential is just (2.4) with $c^2$ replaced by $v^2$. The wave equation for $\mathbf{A}_v$ is (2.7) with a factor of $(1-c^2/v^2)$ multiplying the gradient of the time derivative of the scalar potential. For ease of conception we envision $v > c$, although such a limitation is not real, as the formulas below show.

**A. Gauge function to go from Lorenz gauge to "velocity" gauge**

The construction of the necessary gauge function exactly parallels that of Section III. The scalar potential in the v-gauge is

$$\Phi_v(\mathbf{x},t) = \int d^3x' \frac{1}{R} \rho(\mathbf{x}',t - R/v) \tag{7.2}$$

This means that the time derivative of the gauge function to go from the Lorenz gauge to the v-gauge is

$$\frac{\partial \chi_v}{\partial t} = c \int d^3x' \frac{1}{R} \left[ \rho(\mathbf{x}',t - R/c) - \rho(\mathbf{x}',t - R/v) \right] \tag{7.3}$$

But we can write the square bracket as

$$\left[ \rho(\mathbf{x}',t - R/c) - \rho(\mathbf{x}',t - R/v) \right] = \int_{R/v}^{R/c} d\tau \frac{\partial}{\partial \tau} \rho(\mathbf{x}',t - \tau) \tag{7.4}$$

With the same steps as from (3.5) to (3.6), we find

$$\chi_v(\mathbf{x},t) = -c \int d^3x' \frac{1}{R} \int_{R/v}^{R/c} d\tau\, \rho(\mathbf{x}',t - \tau) \tag{7.5}$$

as the generalization of (3.6).[19]

### B. Vector potential in the v-gauge

In a series of steps completely analogous to those going from above (3.9) to (3.12), we find the gradient of $\chi_v$ to be

$$\chi_v(\mathbf{x},t) = -c \int d^3x' \frac{\hat{\mathbf{R}}}{R^2} \int_{R/v}^{R/c} d\tau\, \tau\, \nabla' \mathbf{J}(\mathbf{x}',t - \tau) \tag{7.6}$$

(B.6) generalizes (3.12). Again paralleling the development from (3.12) to (3.16), we obtain the vector potential in the v-gauge as

$$\mathbf{A}_v(\mathbf{x},t) = \frac{1}{c} \int d^3x' \frac{1}{R} \left[ \mathbf{J}(\mathbf{x}',t - R/c) - \hat{\mathbf{R}}\hat{\mathbf{R}} \cdot \mathbf{J}(\mathbf{x}',t - R/c) + \frac{c^2}{v^2} \hat{\mathbf{R}}\hat{\mathbf{R}} \cdot \mathbf{J}(\mathbf{x}',t - R/v) \right]$$
$$+ c \int d^3x' \frac{1}{R^3} \int_{R/v}^{R/c} d\tau\, \tau \left( 3\hat{\mathbf{R}}\hat{\mathbf{R}} \cdot \mathbf{J}(\mathbf{x}',t - \tau) - \mathbf{J}(\mathbf{x}',t - \tau) \right) \tag{7.7}$$

This result generalizes the Coulomb-gauge (3.16). With $v = c$ the extra terms vanish, leaving the Lorenz gauge potential. For $v \to \infty$ we recover (3.16). Our result (7.7) can be shown with only a small amount of work to be the same as that of Brown and Crothers.[20]

For completeness we record the form of the v-gauge vector potential obtained directly from (7.5), the analog of (3.10) in the Coulomb gauge,

$$\mathbf{A}_v(\mathbf{x},t) = \frac{1}{c} \int d^3x' \frac{1}{R} \left\{ \left[\mathbf{J}(\mathbf{x}',t') - c\hat{\mathbf{R}}\rho(\mathbf{x}',t')\right]_{ret} + \frac{c^2}{v}\hat{\mathbf{R}}\rho(\mathbf{x}',t-R/v) + \frac{c^2\hat{\mathbf{R}}}{R}\int_{R/v}^{R/c} d\tau\, \rho(\mathbf{x}',t-\tau) \right\} \tag{7.8}$$

Here *ret* has the same meaning as in (3.10).

### C. Fields derived from v-gauge potentials

To hammer home the message that the electromagnetic *fields* are properly retarded regardless of the gauge of the potentials, we repeat the exercise of Section V with the v-gauge potentials (7.2) and (7.8). The negative gradient of the v-gauge scalar potential is

$$-\nabla\Phi_v = \int d^3x' \frac{1}{R} \left[ \frac{\hat{\mathbf{R}}}{R}\rho(\mathbf{x}',t') + \hat{\mathbf{R}}\frac{\partial}{v\partial t}\rho(\mathbf{x}',t') \right]_{ret(v)} \tag{7.9}$$

The subscript *ret(v)* means retarded with speed *v*. Correspondingly, *ret(c)* means retardation at the speed of light. The v-gauge vector potential contribution to the electric field is

$$-\frac{\partial}{c\partial t}\mathbf{A}_v(\mathbf{x},t) = \int d^3x' \frac{1}{R} \left\{ \left[ -\frac{\partial \mathbf{J}(\mathbf{x}',t')}{c^2 \partial t} + \hat{\mathbf{R}}\frac{\partial \rho(\mathbf{x}',t')}{c\partial t} \right]_{ret(c)} - \left[\hat{\mathbf{R}}\frac{\partial \rho(\mathbf{x}',t')}{v\partial t}\right]_{ret(v)} - \frac{\hat{\mathbf{R}}}{R}\int_{R/v}^{R/c} d\tau\, \frac{\partial \rho(\mathbf{x}',t-\tau)}{\partial t} \right\}$$

With $\partial\rho/\partial t = -\partial\rho/\partial\tau$ in the integral, after integration the vector potential contribution becomes

$$-\frac{\partial}{c\partial t}\mathbf{A}_v(\mathbf{x},t) = \int d^3x' \frac{1}{R} \left\{ \begin{array}{l} -\frac{\partial \mathbf{J}(\mathbf{x}',t')}{c^2 \partial t} + \hat{\mathbf{R}} \frac{\partial \rho(\mathbf{x}',t')}{c\partial t} + \frac{\hat{\mathbf{R}}}{R}\rho(\mathbf{x}',t') \Bigg|_{ret(c)} \\ -\frac{\hat{\mathbf{R}}}{R}\rho(\mathbf{x}',t') + \hat{\mathbf{R}}\frac{\partial \rho(\mathbf{x}',t')}{v\partial t} \Bigg|_{ret(v)} \end{array} \right. \qquad (7.10)$$

We see that the second line in (7.10), an expression with retardation at the arbitrary speed $v$, just cancels the similarly retarded contribution from the scalar potential (7.9), leaving the electric field **E** equal to the normally retarded result, (5.2). The magnetic field **B** (5.3) follows directly from (7.8) with only the current density term contributing to the curl. We emphasize that the *fields* are causal at speed $c$ despite the arbitrary propagation behavior of the v-gauge potentials (and true whether $v > c$ or $v < c$).

### D. Quasi-static limit of the v-gauge vector potential

In the quasi-static limit retardation is neglected. The time integration in (7.7) can be done. The result is

$$\mathbf{A}_v^{qs}(\mathbf{x},t) = \frac{1}{c} \int d^3x' \frac{1}{R} \left[ \frac{1}{2}(1+\alpha)\mathbf{J}(\mathbf{x}',t) + \frac{1}{2}(1-\alpha)\hat{\mathbf{R}}\hat{\mathbf{R}} \cdot \mathbf{J}(\mathbf{x}',t) \right] \qquad (7.11)$$

where $\alpha = c^2/v^2$. Equation (7.11) is Helmholtz's interpolation between the Neumann and Weber forms of the quasi-static vector potential. It represents a restricted class of different gauges. The Coulomb-gauge result (6.1) comes from $\alpha = 0$ - see (7.1). Note that with the neglect of retardation the scalar potential (7.2) is independent of $v$ and is just the Coulomb-gauge scalar potential for all $v$.

Because (7.1) and (7.11), as well as the gauge function (A.5'), were obtained in 1870 for quasi-static fields by Helmholtz, we should perhaps rechristen the v-gauge as the Helmholtz gauge or class of gauges.

### E. Technical remark

There is a technical point here. Our development of the v-gauge potentials is based on an arbitrary speed of propagation for the scalar potential, either slower or faster than the speed of light, but did not envision an *imaginary speed*. That is what happens for $\alpha = c^2/v^2$ negative. Weber's vector potential requires $\alpha = -1$. Happily, one can show that there are no problems in the quasi-static limit. But detailed consideration of this regime, with its exponentially growing or decaying instead of propagating solutions, would take us too far afield to pursue further here. Baxter[18] points out that $v$ must be real in QED, but a negative $v$ is permitted.

## VIII. HAMILTONIAN OR TEMPORAL GAUGE

The Hamiltonian or temporal gauge is one in which the scalar potential is identically zero. In this case the gauge function to transform from the Lorenz gauge to the temporal gauge has a simple time derivative (from (1.2)),

$$\frac{1}{c}\frac{\partial \chi_T(\mathbf{x},t)}{\partial t} = \Phi_L(\mathbf{x},t) = \int d^3x' \frac{1}{R} \rho(\mathbf{x}',t-R/c) \qquad (8.1)$$

Just as in Section III we integrate both sides with respect to $ct$ to obtain

$$\chi_T(\mathbf{x},t) = c \int d^3x' \frac{1}{R} \int_{t_0}^{t-R/c} dt' \rho(\mathbf{x}',t') \qquad (8.2)$$

Here, in contrast to earlier sections, the lower limit $t_0$ remains. As we see below, its presence is crucial for a description of *electrostatics* in terms of the vector potential alone. Then as before we change variable from $t'$ to $\tau = t - t'$ to get

$$\chi_T(\mathbf{x},t) = c \int d^3x' \frac{1}{R} \int_{R/c}^{t-t_0} d\tau \, \rho(\mathbf{x}',t-\tau) \qquad (8.3)$$

With no scalar potential, all that remains is to calculate the gradient of $\chi_T$:

$$\nabla \chi_T = -\int d^3x' \frac{\hat{\mathbf{R}}}{R} \rho(\mathbf{x}',t-R/c) + \frac{c}{R} \int_{R/c}^{t-t_0} d\tau \, \rho(\mathbf{x}',t-\tau) \qquad (8.4)$$

If we add (8.4) to the Lorenz-gauge vector potential we obtain the vector potential in the temporal gauge to be

$$\mathbf{A}_T(\mathbf{x},t) = \frac{1}{c} \int d^3x' \frac{1}{R} \left[\mathbf{J}(\mathbf{x}',t') - c\hat{\mathbf{R}}\rho(\mathbf{x}',t')\right]_{ret} - \frac{c^2\hat{\mathbf{R}}}{R} \int_{R/c}^{t-t_0} d\tau \, \rho(\mathbf{x}',t-\tau) \qquad (8.5)$$

It takes but a few lines to show that the negative time derivative of (8.5) divided by $c$ gives the familiar retarded electric field (5.2) and its curl, the magnetic field (5.3). The apparent dependence on $(t - t_0)$ cancels out.

To obtain a result involving less of the charge density and more of the current density, we can proceed just as we did in going from (3.9) to (3.11). An integration by parts in (8.4) transforms it into

$$\chi_T = \int d^3x' \frac{\hat{\mathbf{R}}}{R^2} \left\{ -c(t-t_0)\rho(\mathbf{x}',t_0) + c\int_{R/c}^{t-t_0} d\tau\, \tau \frac{\partial \rho(\mathbf{x}',t-\tau)}{\partial \tau} \right\} \quad (8.6)$$

Use of the continuity equation for charge and current as done in going from (3.11) to (3.16) - only the limits of integration are different - and the addition of the Lorenz-gauge vector potential leads to an alternative expression for the temporal-gauge vector potential,

$$\begin{aligned}\mathbf{A}_T = &\frac{1}{c}\int d^3x' \frac{1}{R}\left[\mathbf{J}(\mathbf{x}',t') - \hat{\mathbf{R}}\,\hat{\mathbf{R}}\cdot\mathbf{J}(\mathbf{x}',t')\right]_{ret} - c(t-t_0)\int d^3x' \frac{\hat{\mathbf{R}}}{R^2}\rho(\mathbf{x}',t_0) \\ &- c\int d^3x' \frac{1}{R^3}\int_{R/c}^{t-t_0} d\tau\, \tau\left[3\hat{\mathbf{R}}\,\hat{\mathbf{R}}\cdot\mathbf{J}(\mathbf{x}',t-\tau) - \mathbf{J}(\mathbf{x}',t-\tau)\right]\end{aligned} \quad (8.7)$$

This form of the vector potential is closely similar to the Coulomb-gauge result (3.16). The absence of the scalar potential in this gauge is compensated for by the explicit time dependence on ($t$-$t_0$). Note that for electrostatics, with its vanishing current density and its time-independent charge density, it is the second integral with its coefficient $-c(t-t_0)$ that gives the electrostatic electric field.

## IX. POINCARÉ-MULTIPOLAR-FOCK-SCHWINGER GAUGE

A choice of gauge that goes by the names Poincaré gauge, multipolar gauge, Fock-Schwinger gauge (and other names, too) finds use for atoms and molecules interacting with external electromagnetic fields[21, 22] and also in quantum field theory.[23, 24] Actually, there are two closely related gauges, one constraining the vector potential according to $\mathbf{x}\cdot\mathbf{A} = 0$, usually associated with the name multipolar or Poincaré gauge, and the other its relativistic 4-vector generalization, $x_\mu A^\mu = 0$, usually but not always called the Fock-Schwinger or sometimes the Poincaré gauge.

### A. Nonrelativistic Poincaré or multipolar gauge

The mulitpolar gauge has been discussed in this journal by Kobe[25], Brittin, Rodman Smythe, and Wyss[26], and Skagerstam.[27] Our discussion is therefore mostly brief and derivative. The constraint $\mathbf{x}\cdot\mathbf{A}_M(\mathbf{x},t) = 0$ can be cast in a useful form,

$$\int_0^1 du\, \mathbf{x}\cdot\mathbf{A}_M(u\mathbf{x},t) = 0 \quad (9.1)$$

The integration can be thought of as going along the straight radial line from the origin to the point **x**. With **A** the vector potential in some reference gauge, the relation $\mathbf{A}_M = \mathbf{A} + \nabla \chi_M$ applied to the potentials and gauge function with coordinate $u\mathbf{x}$, (9.1) implies that

$$0 = \int_0^1 du \, \mathbf{x} \cdot \mathbf{A}(u\mathbf{x},t) + \int_0^1 du \, \mathbf{x} \cdot \nabla_{u\mathbf{x}} \chi_M(u\mathbf{x},t) \tag{9.2}$$

Introducing spherical coordinates [$\mathbf{x} = (r, \theta, \phi)$], (9.2) can be written

$$\int_0^1 du \, r \frac{\partial}{\partial(ur)} \chi_M(ur,\theta,\phi,,t) = -\int_0^1 du \, \mathbf{x} \cdot \mathbf{A}(u\mathbf{x},t) \tag{9.3}$$

On the left $r$ is a constant in the integration and $r\partial \chi / \partial(ur) = \partial \chi / \partial u$. The integration is then elementary, with the result that the gauge function is

$$\chi_M(\mathbf{x},t) - \chi_M(0,t) = -\int_0^1 du \, \mathbf{x} \cdot \mathbf{A}(u\mathbf{x},t) \tag{9.4}$$

The vector potential in the multipolar gauge is

$$\mathbf{A}_M(\mathbf{x},t) = \mathbf{A}(\mathbf{x},t) + \nabla \chi_M = \mathbf{A}(\mathbf{x},t) - \nabla \int_0^1 du \, \mathbf{x} \cdot \mathbf{A}(u\mathbf{x},t) \tag{9.5}$$

The gradient acts on $\mathbf{x} \cdot \mathbf{A}(u\mathbf{x},t)$ to give

$$\nabla_\mathbf{x} \mathbf{x} \cdot \mathbf{A}(u\mathbf{x},t) = (\mathbf{x} \cdot \nabla_\mathbf{x}) \mathbf{A}(u\mathbf{x},t) + (\mathbf{A} \cdot \nabla_\mathbf{x})\mathbf{x} + \mathbf{x} \times (\nabla_\mathbf{x} \times \mathbf{A}(u\mathbf{x},t))$$

Introducing spherical coordinates again, and noting that $r\partial/\partial r = (ur)\partial/\partial(ur) = u\partial/\partial u$ in the integral over u, we find

$$\nabla_\mathbf{x} \mathbf{x} \cdot \mathbf{A}(u\mathbf{x},t) = u \frac{\partial}{\partial u} \mathbf{A}(u\mathbf{x},t) + \mathbf{A}(u\mathbf{x},t) + \mathbf{x} \times (\nabla_\mathbf{x} \times \mathbf{A}(u\mathbf{x},t)) \tag{9.6}$$

The first two terms combine to give the derivative with respect to u of ($u\mathbf{A}$). Then in (9.5) the integration over u of the first two terms in (9.6) give $-\mathbf{A}$, which cancels the **A** in (9.5). The net result is that (9.5) becomes

$$\mathbf{A}_M = -\int_0^1 du \, \mathbf{x} \times (\nabla_\mathbf{x} \times \mathbf{A}(u\mathbf{x},t)) = -\mathbf{x} \times \int_0^1 du \, u \, \nabla_{u\mathbf{x}} \times \mathbf{A}(u\mathbf{x},t) \tag{9.7}$$

Finally, with the definition of the magnetic field **B** in terms of **A**, the vector potential in the multipolar gauge is

$$\mathbf{A}_M = -\mathbf{x} \times \int_0^1 du\, u\, \mathbf{B}(u\mathbf{x},t) \tag{9.8}$$

It should not come as a surprise that the scalar potential in the multipolar gauge can be expressed in terms of the electric field. The steps are as follows. The multipolar-gauge scalar potential is

$$\Phi_M = \Phi - \frac{1}{c}\frac{\partial \chi_M}{\partial t} = \int_0^1 du\, \frac{\mathbf{x}}{c} \cdot \frac{\partial}{\partial t}\mathbf{A}(u\mathbf{x},t) - \frac{1}{c}\frac{\partial}{\partial t}\chi_M(0,t) \tag{9.9}$$

With the definition of the electric field in terms of **A** and the integral can written so that

$$\Phi_M = \Phi(\mathbf{x},t) - \int_0^1 du \left[\mathbf{x}\cdot\mathbf{E}(u\mathbf{x},t) + \mathbf{x}\cdot\nabla_{u\mathbf{x}}\Phi(u\mathbf{x},t)\right] - \frac{1}{c}\frac{\partial}{\partial t}\chi_M(0,t)$$

Again with our use of spherical coordinates we have (within the integral over $u$ with $r$ fixed)

$$\mathbf{x}\cdot\nabla_{u\mathbf{x}}\Phi(u\mathbf{x},t) = r\,\partial\Phi/\partial(ur) = \partial\Phi/\partial u$$

The integral over $u$ of that term gives $\Phi(\mathbf{x},t) - \Phi(0,t)$. With its negative sign, the first term cancels the scalar potential in (9.9) and leads to the new scalar potential,

$$\Phi_M = -\int_0^1 du\, \mathbf{x}\cdot\mathbf{E}(u\mathbf{x},t) + \Phi(0,t) - \frac{1}{c}\frac{\partial}{\partial t}\chi_M(0,t)$$

Finally, we choose $\chi_M(0,t)$ so that the two functions of time alone cancel. The scalar potential in the nultipolar gauge is then

$$\Phi_M = -\mathbf{x}\cdot\int_0^1 du\, \mathbf{E}(u\mathbf{x},t) \tag{9.10}$$

Eqs. (9.8) and (9.10) display the potentials in the multipolar gauge in terms of the electric and magnetic fields. We leave it to the reader to verify that the fields calculated from these potentials in the usual way are indeed $\mathbf{E}(\mathbf{x},t)$ and $\mathbf{B}(\mathbf{x},t)$.[28]

The gauge function (9.4) with the choice just made is (up to an additive constant)

$$\chi_M(\mathbf{x},t) = c\int^t dt''\,\Phi(\mathbf{x},t'') - \int_0^1 du\, \mathbf{x}\cdot\mathbf{A}(u\mathbf{x},t) \tag{9.11}$$

Here $\chi_M$ is expressed in terms of the reference potentials. If one wishes an explicit expression in terms of the sources, one may substitute the answers for the reference potentials into (9.11). Stewart[29] has done this with the Lorenz gauge as the reference gauge. If the retarded potentials are written with a time integral over a retarded delta function,

$$d^3x' \frac{1}{R} f(\mathbf{x}',t'=t-R/c) = d^3x' \, dt' \frac{1}{R} f(\mathbf{x}',t')\delta(t'-t+R/c), \qquad (9.12)$$

where f stands for $\rho$ or $\mathbf{J}$, substitution of the Lorenz gauge potentials into (9.11) leads directly to

$$-\int_0^1 du \, \mathbf{x} \cdot \mathbf{A}(u\mathbf{x},t) = -\frac{1}{c} \int d^3x' \, dt' \, \mathbf{x} \cdot \mathbf{J}(\mathbf{x}',t') \int_0^1 du \, \frac{\delta(t'-t+|u\mathbf{x}-\mathbf{x}'|/c)}{|u\mathbf{x}-\mathbf{x}'|} \qquad (9.13)$$

The scalar potential term, which contains an indefinite time integral over the delta function, becomes

$$c \int^t dt'' \, \Phi(\mathbf{x},t'') = -c \int d^3x' \frac{1}{|\mathbf{x}'|} \int_{-}^{|\mathbf{x}'/c|} d\tau \, \rho(\mathbf{x}',t-\tau) \qquad (9.14)$$

The sum of (9.13) and (9.14) give $\chi_M$ explicitly in terms of the sources of charge and current. It bears some semblance to the gauge functions discussed in sections III and VII. There we obtained either a "charge density" expression or a "current density" form. Here the mixture of both types occurs because the multipolar gauge has both Lorenz-gauge potentials transformed into inherently different structures.

**B. Charged-particle interaction Lagrangian with multipolar gauge**

The multipolar gauge or its equivalent is used in quantum mechanical treatments of the interaction of atoms and molecular with external or radiation fields in what amounts to the dipole approximation (or occasionally higher multipoles). We begin classically and consider a number of charged particles in motion in interaction with external electromagnetic fields. For simplicity of notation, let the particles be described formally by continuous charge and current densities,

$$\rho(\mathbf{x},t) = \sum_j \rho_j(\mathbf{x},t) \; ; \quad \mathbf{J}(\mathbf{x},t) = \sum_j \mathbf{J}_j(\mathbf{x},t)$$

The sum is over the individual atoms or molecules. The interaction Lagrangian is then

$$L_{int} = -\int d^3x \, \rho(\mathbf{x},t)\Phi_M(\mathbf{x},t) + \frac{1}{c} \int d^3x \, \mathbf{J}(\mathbf{x},t) \cdot \mathbf{A}_M(\mathbf{x},t) \qquad (9.15)$$

With (9.8) and (9.10) as the potentials we have

$$L_{int} = \int d^3x\ \rho(\mathbf{x},t)\mathbf{x}\cdot\int_0^1 du\ \mathbf{E}(u\mathbf{x},t) - \frac{1}{c}\int d^3x\ \mathbf{J}(\mathbf{x},t)\cdot\mathbf{x}\times\int_0^1 du\ u\ \mathbf{B}(u\mathbf{x},t)$$

If we now assume that the spatial variation of the fields is negligible over the region of support of the charge and current densities of each molecule, we may perform the integrations over $u$ to obtain

$$L_{int} = \sum_j \mathbf{E}(\mathbf{r}_j,t)\cdot\int d^3x\,(\mathbf{x}-\mathbf{r}_j)\rho_j(\mathbf{x},t) + \mathbf{B}(\mathbf{r}_j,t)\cdot\frac{1}{2c}\int d^3x\,(\mathbf{x}-\mathbf{r}_j)\times\mathbf{J}(\mathbf{x},t) \quad (9.16)$$

Here we have assumed that the atoms or molecules are neutral. The coordinates $\mathbf{r}_j$ are the center of mass or center of charge coordinates of each molecule. The integrals in (9.16) are recognized at the definitions of the electric and magnetic dipole moments of the molecules,

$$\mathbf{p}_j(t) = \int d^3x\,(\mathbf{x}-\mathbf{r}_j)\rho_j(\mathbf{x},t)\ ;\quad \mathbf{m}_j(t) = \frac{1}{2c}\int d^3x\,(\mathbf{x}-\mathbf{r}_j)\times\mathbf{J}_j(\mathbf{x},t) \quad (9.17)$$

The interaction Lagrangian then becomes

$$L_{int} = \sum_j \left\{\mathbf{p}_j\cdot\mathbf{E}(\mathbf{r}_j,t) + \mathbf{m}_j\cdot\mathbf{B}(\mathbf{r}_j,t)\right\} \quad (9.18)$$

With the appropriate quantum-mechanical definitions of the dipole (and electromagnetic field) operators, this interaction or the corresponding contribution to the Hamiltonian is employed in atom and molecular calculations,[30, 31, 32] although it is not always obtained by starting with potentials in the multipolar gauge.[33] Often the electric and magnetic fields are expanded around the origin to include their spatial derivatives; quadrupole and higher multipole contributions to (9.18) then emerge.

## C. Fock-Schwinger or relativistic Poincaré gauge

The relativistically invariant gauge condition $x^\mu A_\mu = 0$ can be implemented in the same fashion as the nonrelativistic condition of the preceding section. We just give the obvious results.[27] The gauge function is

$$\chi_{FS}(x) = \chi_0 + \int_0^1 d\zeta\ x^\mu A_\mu(\zeta x) \quad (9.19)$$

where $\chi_0$ is a constant. The 4-vector potential in the Fock-Schwinger gauge is

$$A^{\mu}_{FS}(x) = \int_0^1 d\zeta \; \zeta \; x_\nu \, F^{\nu\mu}(\zeta x) \tag{9.20}$$

The space and time components exhibited separately are

$$\mathbf{A}(\mathbf{x}, x^0) = -\int_0^1 d\zeta \; \zeta \; \left[ x^0 \mathbf{E}(\zeta \mathbf{x}, \zeta t) + \mathbf{x} \times \mathbf{B}(\zeta \mathbf{x}, \zeta t) \right]$$
$$A^0(\mathbf{x}, x^0) = -\int_0^1 d\zeta \; \zeta \; \mathbf{x} \cdot \mathbf{E}(\zeta \mathbf{x}, \zeta t) \tag{9.21}$$

While similar to the multipolar gauge results, the relativistic forms differ in detail.

Skagerstam[27] states a gauge function that connects the covariant gauge $x_\mu A^\mu = 0$ with the multipolar gauge $\mathbf{x} \cdot \mathbf{A} = 0$. Specifically,

$$A^\mu_M = A^\mu_{FS} - \partial^\mu \chi_{Sk} \tag{9.22}$$

where

$$\chi_{Sk}(\mathbf{x}, t) = \int_0^1 d\zeta \int_\zeta^1 d\eta \; t \; \mathbf{x} \cdot \mathbf{E}(\zeta \mathbf{x}, \eta t) \tag{9.23}$$

We leave to the reader verification that this gauge function does the job advertised.[34]

## X. SUMMARY AND CONCLUSIONS

Our purpose is to illustrate explicitly with a number of examples the gauge transformation functions $\chi$ that take one from one set of electromagnetic potentials $(\Phi, \mathbf{A})$ to another $(\Phi', \mathbf{A}')$. In addition, for the Coulomb gauge and the v-gauges, we show explicitly that, whatever the peculiar propagation characteristics of the potentials, the electric and magnetic *fields* derived from them are the same for all choices of gauge.

The choice of gauge is a matter of convenience. The basic connections (1.1) of the fields to the potentials guarantee that the fields are unchanged by a change of gauge (1.2) of the potentials. One might argue that this paper is an empty exercise with obvious foregone conclusions. Yet enough confusion and error continue to appear in the literature that it seems useful to explicate at length with concrete examples. It is also useful, I think, to exhibit the actual gauge functions $\chi$, a thing rarely if ever done in textbooks.

Two features are worthy of note: (i) the relatively simple forms of the vector potential in the Coulomb and v-gauges, with the only added complication (relative to the Lorenz-gauge retarded solutions) of a temporal nonlocality over a finite time interval; and (ii) the little known v-gauge with its scalar potential propagating at an arbitrary speed *v* relative to *c*. The v-gauge illustrates dramatically how arbitrary and "unphysical" the *potentials* can be, yet still yield the same physically sensible *fields*. In the quasi-static limit the v-gauges are equivalent to a class of gauges introduced (without these words) by Helmholtz in 1870.

**ACKNOWLEDGMENTS**

The author thanks Robert N. Cahn for a reading of the manuscript and useful comments. He thanks R. Rosenfelder and D. Dettman for drawing his attention to the prior literature on the v-gauges. The work was supported in part by the Director, Office of Science, Office of High Energy and Nuclear Physics, of the U.S. Department of Energy under Contract DE-AC03-76SF00098.

## APPENDIX: LORENZ-TO-COULOMB GAUGE FUNCTION $\chi_C$ IN TERMS OF THE CURRENT DENSITY

**(a) General result**

To obtain an alternative form in terms of J we consider the heart of (3.6) and perform a (perhaps not obvious) integration by parts in the integration over $d\tau$ by putting $d\tau = -d(R/c - \tau)$:

$$\int_0^{R/c} d\tau \, \rho(\mathbf{x}', t-\tau) = -(R/c - \tau)\rho(\mathbf{x}', t-\tau)\Big|_0^{R/c} + \int_0^{R/c} d\tau \, (R/c - \tau)\frac{\partial}{\partial \tau}\rho(\mathbf{x}', t-\tau)$$

or

$$\int_0^{R/c} d\tau \, \rho(\mathbf{x}', t-\tau) = (R/c)\rho(\mathbf{x}', t) + \int_0^{R/c} d\tau \, (R/c - \tau)\frac{\partial}{\partial \tau}\rho(\mathbf{x}', t-\tau) \qquad (A.1)$$

With the connection between partial derivatives stated above (3.6), we can use the continuity equation for charge and current densities to replace the derivative in (A.1) by $\nabla' \cdot \mathbf{J}(\mathbf{x}', t-\tau)$. If we insert (A.1) into (3.6), the gauge function becomes

$$\chi_C = -\int d^3x' \, \rho(\mathbf{x}', t) - \int d^3x' \int_0^{R/c} d\tau \left(1 - \frac{c\tau}{R}\right) \nabla' \cdot \mathbf{J}(\mathbf{x}', t-\tau) + \chi_0 \qquad (A.2)$$

The first integral is just the total charge Q, assumed to be constant if the current distribution is localized. In the second term we integrate by parts in the $d^3x'$ integration to obtain

$$\chi_C = -Q + \int d^3x' \, \nabla' \int_0^{R/c} d\tau \left(1 - \frac{c\tau}{R}\right) \cdot \mathbf{J}(\mathbf{x}', t-\tau) + \chi_0 \qquad (A.3)$$

Here the gradient operator acts only on the factors of R, not on the current density **J**. It gives zero when acting on the upper limit of the $d\tau$ integration because the integrand vanishes at that upper limit. The gradient does act on the factor $1/R$ to give $\nabla'(1/R) = -\nabla(1/R) = \hat{\mathbf{R}}/R^2$. The result for $\chi_C$ is

$$\chi_C = -c \int d^3x' \frac{1}{R^2} \int_0^{R/c} d\tau \, \tau \, \hat{\mathbf{R}} \cdot \mathbf{J}(\mathbf{x}', t-\tau) \qquad (A.4)$$

We have chosen the irrelevant constant $\chi_0 = Q$ to clear (A.4) of unnecessary quantities. Equation (A.4) is the "current" representation of the gauge function, completely equivalent to the "charge" form, (3.6).

**(b) Helmholtz's quasi-static gauge function**

In 1870 Helmholtz discussed the equivalence of certain different forms of the vector potential[12] and exhibited the scalar function that interpolated between the Neumann form of **A** (now known to be the quasi-static limit of **A** in the Lorenz gauge) to the Weber form (the corresponding limit of **A** in a gauge where $\nabla \cdot \mathbf{A} - \frac{1}{c}\frac{\partial \Phi}{\partial t} = 0$ ).

The quasi-static limit occurs when the source and observation point are such the $R$ is bounded by some limit $R_{max}$ and the fractional change of the current in a time interval $R_{max}/c$ is negligible. Then the change in the current density during the time integration in (A.4) can be neglected. The current density can be replaced by its present value **J**(**x**, *t*) The time integral is then $R^2/2c^2$ and the quasi-static gauge function of Helmholtz is

$$\chi_C^{qs} = -\frac{1}{2c} \int d^3x' \; \hat{\mathbf{R}} \cdot \mathbf{J}(\mathbf{x}',t) \tag{A.5}$$

Actually, (A.5) is a special case of Helmholtz's result, appropriate for the change from Lorenz (Neumann form) to Coulomb gauge. His actual expression is

$$\chi_H^{qs} = -\frac{(1-\alpha)}{2c} \int d^3x' \; \hat{\mathbf{R}} \cdot \mathbf{J}(\mathbf{x}',t) \tag{A.5'}$$

where $\alpha = 1, 0, -1$ correspond to Neumann, Coulomb-gauge, and Weber.

**(c) Alternative route to Helmholtz's quasi-static gauge function**

The result (A.5) can be obtained directly from approximating (3.6), but using slightly more care. It is not sufficient to say that the charge density is unchanged during the time integration - that yields a constant $\chi$. The charge density must be expanded in a Taylor series in time:

$$\chi_C = \chi_0 - c \int d^3x' \frac{1}{R} \int_0^{R/c} d\tau \left[ \rho(\mathbf{x}',t) - \tau \frac{\partial}{\partial t}\rho(\mathbf{x}',t) + \dots \right] \tag{A.6}$$

Integration over $d\tau$ of the first term in the square bracket results in -$Q$, canceling the constant according to our choice above. Integration over $d\tau$ (not *t*) and use of the charge-current continuity equation on the second term yields

$$\chi_C^{qs} = -\frac{1}{2c} \int d^3x' R \; \nabla' \cdot \mathbf{J}(\mathbf{x}',t) \tag{A.7}$$

Now integration by parts casts the gradient operator over onto $R$, with $\nabla' R = -\nabla R = -\hat{\mathbf{R}}$. Thus we obtain (A.5),

$$\chi_C^{qs} = -\frac{1}{2c} \int d^3x' \; \hat{\mathbf{R}} \cdot \mathbf{J}(\mathbf{x}',t) \tag{A.5}$$

___________________________________________

### References and footnotes